\documentclass[a4paper]{jpconf}

\usepackage{graphicx}

\begin{document}

\title{General linear dynamics -- quantum, classical or hybrid}

\author{H-T Elze, G Gambarotta and F Vallone}

\address{Dipartimento di Fisica ``Enrico Fermi'',  
        Largo Pontecorvo 3, I-56127 Pisa, Italia }

\ead{elze@df.unipi.it}

\begin{abstract} We describe our recent proposal of a path integral 
formulation of classical Hamiltonian dynamics. Which leads us here  
to a new attempt at hybrid dynamics, which concerns the direct  
coupling of classical and quantum mechanical degrees of freedom. This 
is of practical as well as of foundational interest and no fully 
satisfactory solution of this problem has been established to date. 
Related aspects will be observed in a general linear ensemble theory, 
which comprises classical and quantum dynamics in the form of Liouville 
and von\,Neumann equations, respectively, as special cases. Considering 
the simplest object characterized by a two-dimensional state-space, 
we illustrate how quantum mechanics is special in several respects 
among possible linear generalizations. 
\end{abstract}

\section{The notorious linearity of the evolution equations}
This article is motivated by discussions of the origin of 
quantum mechanics, as presented, for example, in Refs. \cite{tHooft10,Elze09a}.  
These suggest that quantum mechanics could possibly {\it emerge} by coarse-graining and   
sufficiently far from the Planck scale. Thus, the quantum phenomena would 
arise from dynamics beneath.   
  
Numerous indications exist which support this view. Nevertheless, the general principles 
governing such hypothetical dynamics are under debate  
\cite{Elze09b,Elze08,tHooft07,tHooft06a,tHooft06b,Elze05,Blasone05,Vitiello01,Smolin,Adler,Wetterich08,Isidro08} and, perhaps, have not even been touched upon, so far. 
 
In this situation, it may be worth while to focus on aspects of 
quantum and classical mechanics which bring them as closely together as possible, 
yet sharpen remaining differences. The {\it linearity} of the respective evolution  
equations must certainly be counted among such aspects.

Indeed, it is remarkable that dynamical evolution in physics, to a large extent, can be encoded 
in one or the other of three linear equations: 
\begin{eqnarray}\label{Liouville} 
\partial_t\rho &=&\{ H,\rho\} 
\;\;, \\ [1ex] \label{Schroedinger} 
i\partial_t\Psi &=&\hat H\Psi 
\;\;, \\ [1ex] \label{vonNeumann} 
i\partial_t\hat\rho &=&[\hat H,\hat\rho ] 
\;\;, \end{eqnarray}  
which, respectively, are the {\it Liouville equation} (\ref{Liouville}), governing the 
evolution of a classical phase space density $\rho$ in terms of the 
Poisson bracket $\{\;,\;\}$ and the Hamilton function $H$ of the object under study, 
the {\it Schr\"odinger equation} (\ref{Schroedinger}) for the quantum mechanical wave function(al) 
in terms of the appropriate Hamilton operator $\hat H$, and the {\it von\,Neumann equation}  
(\ref{vonNeumann}) for the density operator $\hat\rho$ in terms of the commutator 
$[\;,\;]$ with $\hat H$.   

We emphasize that all three equations pertain to (or can be interpreted as)  
ensemble theories.  They have the generic structure: 
\begin{equation}\label{linform} 
i\partial_t\;\mbox{``state''}=\hat{\cal L}\;\mbox{``state''} 
\;\;, \end{equation} 
where the ``state'' appears only {\it linearly} in all cases. This first-order in time differential equation, of course,     
assumes widely different forms depending on the forces or dynamics under study.  
That is, the ``Liouville superoperator'' $\hat{\cal L}$ on the right-hand side 
allows to incorporate physics ranging from classical 
non-relativistic single particles all the way to relativistic quantum fields, 
employing a phase space ensemble of initial conditions in classical 
statistical mechanics and the functional Schr\"odinger picture for interacting quantum fields. 

Since the Schr\"odinger equation describes special cases (pure states) 
of situations (pure and mixed states) covered by the von\,Neumann equation,  
we consider Eqs.\,(\ref{Liouville}) and (\ref{vonNeumann}), eventually in 
the form of Eq.\,(\ref{linform}), in the following. 

In passing, we recall that {\it nonlinear} modifications of quantum mechanics 
have been proposed and severely constrained 
in various ways, see, for example, Refs.\,\cite{Iwo,Kibble78,Kibble80,Weinberg}. 

In distinction, modifications of the 
linear Eq.\,(\ref{vonNeumann}), in particular, have been derived  
for ``open systems'' or quantum mechanical objects interacting 
with an environment (such as a heat bath of oscillators, etc.). 
-- Here, one is clearly {\it not} interested in modifying quantum mechanics 
but in applying it to complex  situations. -- 
The generic form of the resulting {\it consistent} dynamics, 
replacing Eq.\,(\ref{vonNeumann}) while preserving the defining properties of 
the density operator, is known as the {\it Lindblad equation} \cite{L1,L2}. 
Its generator provides an important example of a {\it completely positive map}.
Some aspects of this will be discussed in more detail in Section~4, 
in order to contrast it with modifications of the linear dynamics introduced there. 

A hypothetical fundamental coupling of quantum mechanical and classical degrees of freedom 
-- {\it ``hybrid dynamics''} -- instead, 
presents a departure from quantum mechanics. While this does not touch the linearity, 
in the sense of Eq.\,(\ref{linform}), consistency of the resulting theory 
has to be carefully examined.  

We remark that there is also a practical interest in certain forms of hybrid dynamics.  
The Born-Oppenheimer approximation, for example, is based on 
a separation of interacting slow and fast degrees 
of freedom of a compound object. The former are treated as approximately classical 
while the latter as of quantum mechanical nature.~\footnote{Mean field theory, 
based on the expansion of quantum mechanical variables 
into a classical part plus quantum fluctuations, leads to another  
approximation scheme and another form of hybrid dynamics. This is discussed 
more generally for macroscopic quantum phenomena in Ref.\,\cite{Huetal}.} 
It must be emphasized, however, that in these cases, 
hybrid dynamics is considered as relevant {\it approximate description} of  
an intrinsically quantum mechanical object. This has recently been employed, 
for example, in a re-derivation of geometric forces and Berry's phase \cite{ZhangWu06}. 

Such considerations are and will become increasingly important for  
precise manipulations of 
quantum mechanical objects by apparently, for all practical purposes classical means, 
especially in (sub-)nanotechnological devices. 
 
However, this may be also relevant for studies of  
the {\it measurement problem}: Namely, quantum 
theory, endowed with the Copenhagen interpretation, {\it assumes} a  
coupling of the quantum mechanical object to the classical measuring (or manipulating) 
apparatus. However, it remains  
profoundly silent about {\it how} this is realized and how the ensuing state reduction 
(``collapse of the wave function'') proceeds. -- 
Truly hybrid dynamics, which does not arise from an approximation, augmented by effects of environment induced decoherence and anharmonic forces, must have something to say about 
measurement situations, if it can be formulated consistently. 

Presently, we are motivated by interest in the  
{\it back-reaction} effect of quantum fluctuations on classical degrees of freedom, 
in particular if they are physically distinct. We recall here especially  
discussions of the ``semiclassical'' Einstein equation coupling the classical metric 
of spacetime to the expectation value of the energy-momentum tensor of quantized matter 
fields. Can this be made into a consistent hybrid theory leaving gravity unquantized? 
This has recently been re-examined in Ref.\,\cite{Diosi10}; earlier related work includes  
Refs.\,\cite{Diosi84,DiosiRev05,Penrose98,Adler03,Mavromatosetal92,Pullinetal08,Hu09,Diosi09}. 
 
Concerning the origin of quantum mechanics from a coarse-grained  
deterministic dynamics \cite{tHooft10,Elze09a}, to which we 
alluded above, the back-reaction problem can be more provocatively stated 
as the problem of the interplay of fluctuations among underlying deterministic and 
emergent quantum mechanical degrees of freedom. Or, in short:  
{\it ``Can quantum mechanics be seeded?''}

In this paper, we present a few ingredients which may be helpful in further 
studies of these questions. -- 
In the following Section~2, we begin with a brief recapitulation of 
our path integral formulation of classical Hamiltonian dynamics.  
Some simple observations following from this lead us to  
take up the subject of hybrid dynamics in Section~3. 
In Section~4, we continue the discussion from a different angle, 
namely with some remarks on problems encountered when trying to embed 
classical or quantum mechanics in a more general linear dynamics.  

\section{A path integral for classical Hamiltonian dynamics} 
Our aim here is to solve the Liouville equation (\ref{Liouville}) with the help of a suitable 
propagator which, in turn, is represented as a path integral. We summarize essential 
steps, while more details of the derivation can be found in Ref.\,\cite{EGV10}. 

The following considerations are rather independent of the number of degrees of freedom and  
apply to matrix or Grassmann valued variables as well; field theories require a classical 
functional formalism, which has been considered elsewhere~\cite{Elze05,Elze07}. 
We consider a one-dimensional system, for simplicity.

\subsection{The quantum-like version of the Liouville equation}
We assume conservative forces acting on the classical object under study 
and that Hamilton's equations are determined by the generic Hamiltonian function:  
\begin{equation}\label{HamiltonianF} 
H(x,p):=\frac{1}{2}p^2+V(x) 
\;\;, \end{equation} 
where $x$ and $p$ denote generalized coordinate and momentum, respectively, and where 
$V(x)$ stands for an external potential (a mass parameter will be inserted later). 

An ensemble of such objects can be described by 
a probability distribution function $\rho$ depending on the $x,p$-coordinates of 
phase space and time. 
Such a distribution evolves according to the {\it Liouville equation}: 
\begin{equation}\label{LiouvilleEq} 
-\partial_t\rho =\frac{\partial H}{\partial p}\cdot\frac{\partial \rho}{\partial_x}
-\frac{\partial H}{\partial x}\cdot\frac{\partial \rho}{\partial_p}
=\big\{p\partial_x-V'(x)\partial_p\big\}\rho  
\;\;, \end{equation} 
with $V'(x):=\mbox{d}V(x)/\mbox{d}x$. -- The relative minus sign 
in the Poisson bracket, or between terms here,   
reflects the symplectic phase space symmetry. It will give rise to a 
commutator structure, which reminds one of quantum mechanics, as we shall see momentarily.   

Following a Fourier transformation, $\rho (x,p;t)=\int\mbox{d}y\;\mbox{e}^{-ipy}\rho (x,y;t)$, 
and a transformation of the effectively {\it doubled number of spacelike coordinates},  
\begin{equation}\label{coordtrans} 
Q:=x+y/2\;\;,\;\;\;q:=x-y/2  
\;\;, \end{equation} 
the Liouville equation becomes: 
\begin{eqnarray}\label{Schroed} 
i\partial_t\rho &=&\big\{ \hat H_Q-\hat H_q+{\cal E}(Q,q)\big\}\rho 
\;\;, \\ [1ex] \label{HX} 
\hat H_\chi &:=&-\frac{1}{2}\partial_\chi ^{\;2}+V(\chi )\;\;, 
\;\;\;\mbox{for}\;\;\chi =Q,q 
\;\;, \\ [1ex] \label{I} 
{\cal E}(Q,q)&:=&(Q-q)V'(\frac{Q+q}{2})
-V(Q)+V(q)\;=\;-{\cal E}(q,Q)
\;\;. \end{eqnarray}  
Thus, it bears strong resemblance to the {\it von\,Neumann equation}, cf. 
Eq.\,(\ref{vonNeumann}),  
considering $\rho (Q,q;t)$ as matrix elements of a density operator $\hat \rho (t)$. 

We automatically recover the Hamiltonian operator $\hat H$  
related to the Hamiltonian function, Eq.\,(\ref{HamiltonianF}),  
as in quantum theory. Furthermore, reality and normalization of the phase space 
probability distribution $\rho (x,p;t)$ translate into hermiticity and trace  
normalization of the density operator $\hat \rho (t)$ \cite{EGV10}. 

However, an essential dynamical feature here consists   
in the interaction ${\cal E}$ between 
{\it bra-} and {\it ket-}states. Thus, generally, 
the Hilbert space and its dual are coupled by 
a genuine {\it superoperator}, a concept to be defined 
in the following 
subsection.~\footnote{The interaction ${\cal E}$ is antisymmetric under 
$Q\leftrightarrow q$. It follows that the complete   
Liouville superoperator on the right-hand side of Eq.\,(\ref{Schroed}), to be compared 
with Eq.\,(\ref{linform}), has a symmetric spectrum 
with respect to zero and, in general, will not be bounded below. 
Related observations were discussed, for example, in  
Refs.~\cite{tHooft06a,Elze05,Blasone05,Vitiello01}.}  


It is remarkable that the interaction between 
bra- and ket-states vanishes under certain 
circumstances: 
\begin{equation}\label{Ezero}
{\cal E}\equiv 0\;\;\Longleftrightarrow\;\;
\mbox{potential}\;V(x)\;\mbox{is constant, linear, or harmonic} 
\;, \end{equation} 
rendering a Liouville superoperator of quantum mechanical form, 
i.e., as in 
Eq.\,(\ref{vonNeumann}).~\footnote{Analogously, the vanishing of ${\cal E}$ in 
a field theory amounts to having massive or massless free fields, with or without 
external sources, and with or without bilinear couplings. In these cases, anharmonic 
forces or interactions are absent.} 
This has been discussed in Ref.\,\cite{Elze09a} under the perspective of having 
quantum phenomena emerge due to discrete spacetime structure. 

In the following, we will study in more detail 
the classical dynamics described by Eq.\,(\ref{Schroed}), or by  
appropriate generalizations, and pay particular attention to the presence of the 
superoperator ${\cal E}$, when comparing with the von\,Neumann equation 
and its solution by a propagator.  

\subsection{The superspace}
The dynamics of density operators in the general form of Eq.\,(\ref{linform}) can 
be conveniently rewritten by introducing the concept of {\it superspace}, also called   
{\it Liouville space} \cite{EGV10,superspace}.~\footnote{No relation with the case of 
supersymmetry is implied.}     

Let $\hat{H}$ denote a Hamiltonian operator, as in quantum 
theory, and let $\{ |j\rangle\}$, $j=1,\dots ,N$, present a complete orthonormal 
set of basis states, 
assuming here that the relevant Hilbert space is $N$-dimensional. 
Then, the matrix elements of the von\,Neumann equation (\ref{vonNeumann}) read:  
\begin{equation}\label{vN}
i \partial_t \rho_{jk} = [(\hat{H}\hat{\rho})_{jk}- (\hat{\rho}\hat{H})_{jk}]\;\;, \;\;\;  
j,k = 1,2,\dots, N
\;\;, \end{equation}
with a density matrix $\rho$ of $N^2$ elements. Or, written as in Eq.\,(\ref{linform}):   
\begin{equation}\label{vN1}
i \partial_t \rho_{jk} = \sum_{l,m} {\cal L}_{jk,lm} \rho_{lm}\equiv
\big (\hat{\cal L}\hat\rho\big )_{jk} 
\;\;, \end{equation}
where the {\it Liouville superoperator} $\hat {\cal L}$ is now defined by its matrix elements: 
\begin{equation}\label{LvN}
{\cal L}_{jk,lm}:= H_{jl} \delta_{mk} - \delta_{jl}H_{mk} 
\;\;. \end{equation}
This simple rewriting may suggest to introduce a space in which the 
density operator is 
a vector. This is the role of the Liouville space (or superspace). 

The dynamics of density operators  
can then be described in parallel for classical and quantum mechanics. In general, they will 
differ, of course, by the precise form of the superoperator, examples of which we have seen 
here and in the previous subsection.    
 
Given the Hilbert space, as above, the density operator can be expanded as:  
\begin{equation}
\hat{\rho}= \sum_{j,k} \rho_{jk} |j \rangle \langle k| 
\;\;. \end{equation}
The family of $N^2$ operators $|j\rangle \langle k|$, with $j,k=1,\cdots,N$, can 
be interpreted as a complete set of matrices, or vectors, such that the density operator 
becomes: 
\begin{equation}
|\rho \rangle= \sum_{j,k} \rho_{jk} |jk \rangle \rangle 
\;\;, \end{equation}
where the ``ket'' $|jk \rangle \rangle$ denotes the Liouville space {\it vector} representing the 
Hilbert space {\it operator} $|j\rangle \langle k|$. Similarly, we introduce a ``bra''-vector
$ \langle \langle jk|$ as the Hermitian conjugate of $|jk \rangle \rangle$.

Consequently, any operator $\hat{A}$ is represented by a vector, denoted by 
$|A \rangle \rangle $, and can be expanded as:  
\begin{equation} \label{opex}
| A \rangle \rangle= \sum_{j,k} A_{jk} |jk \rangle \rangle
\;\;, \end{equation}
where $A_{jk}$ are the usual matrix elements $\langle j| \hat{A}| k \rangle$. -- 
Furthermore, with a bra-vector $ \langle \langle B |$ representing $\hat{B}^{\dag}$, 
the scalar product of two operators is defined by: 
\begin{equation}
\langle \langle B| A \rangle \rangle :=\mbox{Tr} (\hat{B}^{\dag}\hat{A})   
\;\;. \end{equation}
This implies the orthonormality condition: 
\begin{equation}\label{orthogonality} 
\langle \langle jk|mn \rangle \rangle =\mbox{Tr}(|k\rangle \langle j| m \rangle \langle n|)=
\delta_{kn} \delta_{jm} 
\;\;, \end{equation} 
in analogy to and based on $\langle j|k\rangle = \delta_{jk}$. -- Finally, consider the scalar product: 
\begin{equation}
\langle \langle jk| A \rangle \rangle =\mbox{Tr}(|k \rangle \langle j| \hat{A})=
\sum_{l} \langle l|k \rangle \langle j |\hat{A}| l \rangle= \langle j| \hat{A}|k \rangle \equiv A_{jk}
\;\;. \end{equation} 
Upon substitution in Eq.\,(\ref{opex}), this yields: 
\begin{equation}
|A \rangle \rangle = \sum_{j,k} |jk \rangle \rangle \langle \langle jk | A \rangle \rangle 
\;\;. \end{equation}
This is consistent with the following completeness relation in Liouville space: 
\begin{equation}\label{completeness}
\sum_{j,k} |jk \rangle \rangle \langle \langle jk | = {\mathbf 1} 
\;\;. \end{equation}
It is easy to see now that the Liouville
space is a linear space and that the density operator $\hat{\rho}$, in particular, is a 
vector in this space. -- To conclude these formal considerations, we define a linear operator 
in terms of its matrix elements: 
\begin{equation}
\hat{\cal F}:= \sum_{j,k,m,n} |jk \rangle \rangle \langle \langle jk |
\hat{\cal F}|mn \rangle \rangle \langle \langle mn| \equiv   
\sum_{j,k,m,n}{\cal F}_{jk,mn}|jk \rangle \rangle \langle \langle mn| 
\;\;, \end{equation} 
an example of which is given in Eq.\,(\ref{LvN}). 

The importance of Liouville space for classical {\it and} quantum dynamics resides in that 
Liouville and von\,Neumann equations, both, assume the form of Eq.\,(\ref{linform}), 
incorporating an appropriate superoperator $\hat {\cal L}$, cf. 
Eqs.\,(\ref{Schroed})--(\ref{I}) and (\ref{vN})--({\ref{LvN}), respectively. 
Thus, we find a close formal similarity between the structure of these equations 
and the Schr\"odinger equation (\ref{Schroedinger}).

Therefore, it is plausible that formal derivations or techniques concerning the solution of 
the Schr\"odinger equation can be transferred to the cases of Liouville or 
von\,Neumann equations with the help of Liouville space notions. This regards  
perturbation theory as much as nonperturbative methods, the path integral approach  
in particular, to which we turn next.  

\subsection{The Liouville propagator and path integral}
The technical ingredients of the Feynman path integral approach, 
for the derivation of quantum mechanical propagators in particular, are very 
well known \cite{Schulman,Kleinert}. We will make use of these ideas, in order 
to derive a path integral for the propagator of density matrices based on 
the Liouville space formulation of the preceding subsection. 

Our derivation relies on the close formal similarity between the classical 
Liouville equation and the von\,Neumann equation on one hand side and the 
Schr\"odinger equation on the other, in the appropriate representation that  
we discussed.~\footnote{In this subsection, we reinstate $\hbar$ explicitly.}   

In particular, the formal solutions of the classical Liouville equation 
and of the quantum mechanical von\,Neumann equation, both, can be written in the form: 
\begin{equation}
|\rho(t)\rangle\rangle =\mbox{e}^{-i\hat {\cal L}(t-t_0)/\hbar}|\rho(t_0)\rangle\rangle 
\;\;, \end{equation} 
where $\hat {\cal L}$ is the relevant Liouville superoperator. Here, we have: 
\begin{equation}
\langle\langle Q,q|\hat {\cal L}|Q',q'\rangle\rangle =
\delta (Q-Q')\delta (q-q')\big (\hat H(Q)-\hat H(q)+{\cal E}(Q,q))
\;\;, \end{equation} 
where $\hat H$ denotes the appropriate Hamilton operator in coordinate representation, as 
indicated, which alone is relevant for the von\,Neumann equation, while ${\cal E}$ 
represents the additionally present superoperator term for classical dynamics, cf. 
Subsection\,2.1. 

In order to solve the problem of time evolution in the present case, we need to know 
the (super)matrix elements entering the propagation equation: 
\begin{equation}\label{rhoprop}
\langle\langle Q,q|\rho(t)\rangle\rangle =\int\mbox{d}Q'\mbox{d}q'\;
\langle\langle Q,q|\mbox{e}^{-i\hat {\cal L}(t-t_0)/\hbar}|Q',q'\rangle\rangle\langle\langle Q',q'|\rho(t_0)\rangle\rangle
\;\;, \end{equation} 
in analogy to the case of a state vector evolving according the Schr\"odinger equation. 
Not surprisingly, we can now follow the usual steps \cite{Schulman,Kleinert}, in order 
to construct the path integral representation of our propagator \cite{EGV10}. 

In this derivation, one has to pay attention to a suitable generalization of the 
Trotter product formula. This works out in a straightforward way; the relevant 
definitions and details of the proof are given in Ref.\,\cite{FabThesis}.  

Rewriting the Eq.\,(\ref{rhoprop}) as: 
\begin{equation}\label{rhoprop1} 
\rho (Q,q;t) = \int\mbox{d}Q'\mbox{d}q'\;{\cal G}(Q,q;t|Q',q';t_0)\rho (Q',q';t_0) 
\;\;, \end{equation} 
our interest is to know the {\it superpropagator} ${\cal G}$. 

Following the recipe to arrive at a path integral representation,  
we make use here of suitably inserted complete sets of superspace vectors, such as: 
\begin{equation}\label{supercomplete} 
\int\mbox{d}Q\mbox{d}q\; |Q,q\rangle\rangle\langle\langle Q,q|={\mathbf 1} 
\;\;, \end{equation} 
and, correspondingly, for momentum space, cf. Eq.\,(\ref{completeness}).  
Using the plane wave relation between coordinate and momentum eigenfunctions, we also employ: 
\begin{equation}\label{planewaves}
\langle\langle P,p|=\frac{1}{2\pi\hbar }\int\mbox{d}Q\mbox{d}q\;
\exp\big (-\frac{i}{\hbar}(PQ-pq)\big )\langle\langle Q,q| 
\;\;. \end{equation} 
Furthermore, the orthogonality relation 
$\langle\langle Q,q|Q',q'\rangle\rangle =\delta (Q-Q')\delta(q-q')$, cf. 
Eq.\,(\ref{orthogonality}), implies: 
\begin{equation}\label{transf} 
\langle\langle P,p|Q,q\rangle\rangle =\frac{1}{2\pi\hbar }
\exp\big (-\frac{i}{\hbar}(PQ-pq)\big )
\;\;. \end{equation} 
Then, with all following steps of the derivation in parallel with the usual ones, it is 
straightforward to obtain the {\it Liouville path integral}~\cite{EGV10,FabThesis}: 
\begin{equation}\label{superpropagator} 
{\cal G}(Q_f,q_f;t|Q_i,q_i;t_0)=\int {\cal D}Q{\cal D}q
\;\exp\big (\frac{i}{\hbar}{\cal S}[\dot Q,Q;\dot q,q]\big ) 
\;\;, \end{equation} 
with the boundary conditions $Q(t_i)=Q_i$, $q(t_i)=q_i$, $Q(t_f)=Q_f$, and $q(t_f)=q_f$, 
otherwise unrestricted paths, and with the {\it superaction} ${\cal S}$ defined by: 
\begin{equation}  
\label{superaction1} 
{\cal S}:=\int_{t_0}^t\mbox{d}\tau\;\Big ({\textstyle \frac{m}{2}}\dot Q^2-V(Q)
-\big ({\textstyle \frac{m}{2}}\dot q^2-V(q)\big )
-{\cal E}(Q,q)\Big )
\;\;, \end{equation} 
for a particle of mass $m$. 
We recall that ${\cal E}\equiv 0$ corresponds to 
evolution according to the von\,Neumann equation, whereas ${\cal E}\neq 0$ represents 
the {\it only} modification  due to classical dynamics, in accordance with the Liouville 
equation, cf. Eqs.\,(\ref{Schroed})--(\ref{I}). 

However simple this result may seem, the Eqs.\,(\ref{superpropagator})--(\ref{superaction1}) 
describe time evolution of the full density matrix. The 
particular new feature is that formally {\it classical dynamics is treated on the same   
footing as quantum mechanics}, differing only in the action entering the phase in the 
integrand of the path integral.

These considerations are limited neither by one dimension nor by   
single-particle physics, but can be extended all the way to relativistic 
field theories. This offers new calculational tools, new approximation methods 
in particular, and may be of interest for applications in classical statistical mechanics.  
In the following, however, we turn to the quantum-classical divide. 

\subsection{Simple properties of the Liouville path integral}
First of all, it seems worth while to record a few simple properties of the 
Liouvile path integral.   

We observe that in the absence of forces, $V\equiv 0$, the classical and quantum 
mechanical propagators {\it coincide}, according to 
Eqs.\,(\ref{superpropagator})--(\ref{superaction1}). Which implies that classical 
and quantum mechanical behaviour can differ at most in the initial states that 
are being propagated, in this case. -- This holds true even for $V\neq 0$, 
if ${\cal E}\equiv 0$, i.e., for harmonic forces, cf. (\ref{Ezero}). -- Thus, 
the cherished ``textbook effect'' of the {\it spreading of  
wave packets} is not a peculiar (kinematical) quantum  effect, but rather an effect 
of the particular states considered! 

A simple calculation, taking into account our above transformations from 
$x,p$- to $Q,q$-coordinates (and back), shows that a {\it classical point particle} with 
initial phase space distribution $\propto\delta (x-x_0)\delta (p-p_0)$ is 
propagated to the distribution $\propto\delta (x-x_0-p_0t/m))\delta (p-p_0)$, 
i.e. along a straight line path, as expected. -- Instead, a free  
Gaussian wave packet in $Q,q$-space spreads in the way described in textbooks on quantum mechanics.  

Finally, we can convince ourselves that the classical dynamics is properly 
represented by the Liouville propagator, in general, by undoing the coordinate 
transformations (\ref{coordtrans}) directly in the path integral representation of 
the superpropagator: 
\begin{equation}\label{coordtransInverse} 
y:=Q-q\;\;,\;\;\;x:=\frac{Q+q}{2}   
\;\;, \end{equation} 
a transformation with unit Jacobian. Following a partial integration of  
the superaction, the path integral over the $y$-coordinate simply yields a 
functional $\delta$-``function'' and results in: 
\begin{equation}\label{superpropagator1} 
{\cal G}(x_f,y_f;t|x_i,y_i;t_0)=\int {\cal D}x\;  
\mbox{exp}(im\dot x_fy_f)\;\delta [m\ddot x+V'(x)]\;\mbox{exp}(-im\dot x_iy_i) 
\;\;, \end{equation} 
with the boundary conditions $x(t_{i,f})=x_{i,f}$, similarly for the 
velocities, and where $y_{i,f}$ 
denote the initial and final values of the $y$-variable, respectively.  
The phase factors stem from the partial integration of the superaction. 

Thus, we find the expected result that only paths following solutions of the 
{\it classical} equations of motion contribute to the propagator. The role of the 
phase factors is easily understood by recalling Eq.\,(\ref{rhoprop1}):  
After the inverse coordinate transformations (\ref{coordtransInverse}), one 
of the two ordinary integrations there becomes here, including the relevant (initial) phase 
factor: 
\begin{equation}\label{phaseintegral} 
\int\mbox{d}y_i\;\mbox{e}^{-im\dot x_iy_i}\rho (x_i,y_i;t_0)=
\rho (x_i,m\dot x_i\equiv p_i;t_0) 
\;\;, \end{equation} 
i.e., this incorporates the inverse Fourier transformation, back to the momentum variable 
of phase space. The second (final) phase factor, then, is necessary for  
the propagator to fulfil the important semi-group property \cite{Schulman}.  

\section{Hybrid dynamics for two interacting objects} 
We have seen that the path integral for the propagator of the classical Liouville equation 
shows significant similarity with the propagator of the quantum mechanical 
von\,Neumann equation. This suggests a new perspective on {\it hybrid dynamics} --  
the hypothetical direct coupling of quantum and classical degrees of freedom that we discussed 
in Section\,1.   

In particular, we are interested here in the propagator 
for the density matrix of a bi-partite system, composed of a classical and a quantum mechanical 
particle of masses $m,m'$ in external potentials $V,v$, respectively. 
Correspondingly, the relevant coordinates 
will be denoted by $Q,q$ and $Q',q'$, respectively. Our previous 
considerations lead us to propose the following   
{\it hybrid superpropagator} (cf. Eq.\,(\ref{superpropagator})): 
\begin{equation}\label{hsuperpropagator} 
{\cal G}_{\mbox{h}}(Q_f,q_f;Q'_f,q'_f;t|Q_i,q_i;Q'_i,q'_i;t_0)=
\int {\cal D}Q{\cal D}q{\cal D}Q'{\cal D}q'
\;\exp\big (\frac{i}{\hbar}{\cal S}[\dot Q,Q;\dot q,q;\dot Q',Q';\dot q',q']\big ) 
\;\;, \end{equation} 
where the superaction naturally consists of three contributions, 
${\cal S}\equiv {\cal S}_{\mbox{cl}}+{\cal S}_{\mbox{qm}}+{\cal S}_{\mbox{h}}$: 
\begin{eqnarray}\label{Scl}  
{\cal S}_{\mbox{cl}}&:=&\int_{t_0}^t\mbox{d}\tau\;\Big (
{\textstyle \frac{m}{2}}\big (\dot Q^2-\dot q^2\big )-(Q-q)V'(\frac{Q+q}{2}) 
\Big )
\;\;, \\ [1ex] \label{Sqm}
{\cal S}_{\mbox{qm}}&:=&\int_{t_0}^t\mbox{d}\tau\;\Big (
{\textstyle \frac{m}{2}}\big (\dot Q'^2-\dot q'^2\big )
-\big (v(Q')-v(q')\big ) 
\Big )
\;\;, \\ [1ex] \label{Sh}
{\cal S}_{\mbox{h}}&:=&-\int_{t_0}^t\mbox{d}\tau\; {\cal V}(Q,q;Q',q') 
\;\;. \end{eqnarray} 
Several remarks are in order here. The contribution ${\cal S}_{\mbox{cl}}$ is the 
classical superaction arrived at in Eqs.\,(\ref{superpropagator})--(\ref{superaction1}); 
we just inserted ${\cal E}$, Eq.\,(\ref{I}), explicitly and collected terms. Correspondingly,   
the contribution ${\cal S}_{\mbox{qm}}$ is the action for the propagator of the 
von\,Neumann equation, i.e., it describes the quantum mechanical particle in the usual way. 
Thus, in the absence 
of the interaction term ${\cal S}_{\mbox{h}}$, the hybrid superpropagator 
${\cal G}_{\mbox{h}}$ factorizes into the corresponding ones for Liouville and 
von\,Neumann equations. This would consistently describe a composite system  
of two independent, classical and quantum mechanical particles.  

The interaction term ${\cal S}_{\mbox{h}}$ introduces the coupling responsible for 
{\it hybrid dynamics}. -- For our present purposes, we restrict this 
coupling by two consistency requirements:~\footnote{The question of 
consistency of the map induced by the superpropagator which evolves the quantum-classical 
hybrid density will be discussed in more detail elsewhere.} \\ ({\bf A}) 
${\cal S}_{\mbox{cl}}+{\cal S}_{\mbox{h}}$ describes a classical particle 
interacting with another particle (coordinates $Q',q'$), as if the latter were 
{\it classical}; \\ ({\bf B})     
${\cal S}_{\mbox{qm}}+{\cal S}_{\mbox{h}}$describes a quantum mechanical particle 
interacting with another particle (coordinates $Q,q$), as if the latter were 
{\it quantum mechanical}. 

In view of 
the general form of the external potential terms in Eqs.\,(\ref{Scl}) and (\ref{Sqm}),   
these conditions can only be fulfilled, if ${\cal V}(Q,q;Q',q')$ is {\it harmonic}, 
i.e., is a polynomial of degree less than or equal to two in all variables. Thus, 
the hybrid coupling is of a form that admits a classical or quantum mechanical 
interpretation, depending on whether it is viewed from the classical or quantum mechanical 
subsystem. 

An example is provided by a distance dependent oscillator potential, 
$\lambda^{-1}{\cal V}:=(x-x')^2$. Following the derivation of the classical superpropagator, 
this becomes, in the above coordinates: 
\begin{eqnarray}
\lambda^{-1}{\cal V}(Q,q;Q',q')&=&
\left ((Q-q)\partial_{(Q+q)/2}+(Q'-q')\partial_{(Q'+q')/2}\right )
\left ((Q+q)/2-(Q'+q')/2\right )^2
\nonumber \\ [1ex] \label{CLQM}
&=&(Q-q)(Q+q)+(Q'-q')(Q'+q')^2-2(QQ'-qq')
\;\;, \end{eqnarray} 
i.e., it can be seen as coupling between two classical particles, fulfilling condition ({\bf A}). 
On the other hand, the separable oscillator terms here also  
equal $Q^2-q^2+Q'^2-q'^2$, i.e., are of quantum mechanical form, in accordance with 
(\ref{Ezero}), and the bilinear coupling can as well be seen as quantum mechanical, fulfilling condition 
({\bf B}). 

Interesting consequences of these conditions, determining a {\it harmonic hybrid interaction},  
are summarized in the following Table: 
\\  
\begin{center}
\begin{tabular}{|lll|l|}
\hline    
$V$&$v$&${\cal V}$&resulting dynamics \\
\hline 
h&h&h&CL or QM \\ 
anh&h&h&CL \\ 
h&anh&h&QM \\ 
anh&anh&h&{\bf ???} \\ 
\hline  
\end{tabular}  
\end{center}  
{\small \vskip 0.2cm Indicated are the 
nature of the potentials acting on the classical and quantum mechanical 
particles and their interaction, $V$, $v$, and ${\cal V}$, respectively -- h: harmonic; anh: anharmonic -- and the character of the 
resulting hybrid dynamics -- CL: classical; QM: quantum mechanical.}  
\vskip 0.25cm
For example, if 
the classical and quantum mechanical particles are governed by a harmonic potential and 
an anharmonic potential, respectively, then the resulting dynamics of the 
composite system can be considered as that of two bilinearly coupled quantum mechanical 
particles, one of which moves in a harmonic potential. -- The case of both potentials 
being anharmonic is most interesting. However, no general statements can be made in 
this case without further study. 
The quantum mechanical or classical behaviour of the 
composite system might very well depend on {\it where} (concerning its variables) one 
is looking at this object!~\footnote{These findings might help with  
the problem posed in Section~1: {\it ``Can quantum mechanics be seeded?''}}   

\subsection{Intra- and interspace entanglement}
There is a {\it qualitative difference between CL and QM}, contained in 
the full path integral for the superpropagator, 
Eqs.\,(\ref{hsuperpropagator})--(\ref{Sh}). 
Following standard arguments which lead back from a path integral  
to an equivalent equation of motion for the (hybrid) density operator $\hat\rho$, 
several related observations may be interesting.  

The QM evolution is 
generated by a commutator of the Hamiltonian with the density operator. Generally, this  
superposes and, for multi-partite systems, in particular, entangles underlying 
bra- and ket-states separately, $\propto H_{ij}\rho_{jk}-\rho_{ij}H_{jk}$ 
(using a convenient notation with discrete indices). For a bi-partite system, 
it is revealing to write the relevant interaction terms explicitly: 
\begin{equation}\label{QMentangle}
[\hat H_{int},\hat\rho ]=\hat H_1\hat\rho_1\otimes\hat H_2\hat\rho_2-\hat\rho_1\hat H_1 
\otimes\hat\rho_2\hat H_2 
\;\;, \end{equation} 
for an interaction $\hat H_{int}:=\hat H_1\otimes\hat H_2$, with the factors acting on 
subsystems ``1'' and ``2'', respectively, and where $\hat\rho =\hat\rho_1\otimes\hat\rho_2$,   
for a separable initial state. 
This has been called {\it dynamically assisted entanglement generation}, see, for example, Refs.\,~\cite{Jacquod,JacquodRev,Hornberger}. 

It may come as a surprise that the CL evolution does this just as well, 
due to the structure of the superoperator. For polynomial interactions, 
in particular, the superoperator 
{\it always} contains a contribution proportional to the usual QM terms. 

However, the CL evolution, generally, produces additional correlations in $\hat\rho$, 
due to terms contained in ${\cal L}_{ij;kl}\rho_{kl}$  
which {\it entangle bra- and ket-states}. -- 
In comparison with Eq.\,(\ref{QMentangle}), such terms can have the unfamiliar structure: 
\begin{equation}\label{CLentangle} 
\hat H'_1\hat\rho_1\otimes\hat\rho_2\hat H'_2-\hat\rho_1\hat H'_1\otimes\hat H'_2\rho_2 
\;\;, \end{equation} 
which differs decidedly from a commutator. -- This leads us to distinguish {\it intra-} 
(i.e., within given tensor product Hilbert space of subsystems ``1'' and ``2'') and 
{\it inter-space entanglement} (i.e., between said Hilbert space and its dual).  

Consider, for example, the anharmonic potential 
$V(x_1-x_2):=\lambda (x_1-x_2)^4$ for a bi-partite system consisting of particles ``1'' 
and ``2''. Similarly as before, this leads here to the interaction: 
\begin{equation}\label{2particle} 
{\cal V}(Q_1,q_1;Q_2,q_2)=
\frac{1}{2}\lambda\big (Q_1-q_1-(Q_2-q_2)\big )\big (Q_1+q_1-(Q_2+q_2)\big )^3 
\;\;, \end{equation} 
in terms of, by now, familiar variables, taking into account subsystems ``1'' and ``2'' with  
$Q$'s and $q$'s refering to bra- and ket-states, respectively. Besides separable terms, 
$\propto (Q_a-q_a)(Q_a+q_a)^3,\; a=1,2$, there are terms which mix (and entangle) 
variables of both subsystems, as usual in QM. However, there are also additional 
terms that refer to Hilbert space and its dual simultaneously (and entangle corresponding 
states), for example, $\propto Q_aQ_bq_b^2,\; b\neq a$.   

In retrospect, somehow, such difference between CL and QM 
evolution had to be expected:  
instead with superstates $|Q,q\rangle\rangle$, we could have 
worked with superstates $|x,p\rangle\rangle$, relating to coordinates and momenta of 
the classical theory. There, coordinates and momenta end up tightly correlated, due to 
Hamilton's equations, and produce inter-space entanglement in an interacting bi-partite 
system. 

Thus, the confrontation of CL with QM, as in our side-by-side study,   
is quite revealing. In particular, we speculate that this opens new views on  
generating entanglement in multipartite systems, perhaps, by evolving through 
quasiclassical stages or by making use of decohered intermediary 
states.~\footnote{Previous considerations of the semiclassical regime, such as in 
Refs.~\cite{Jacquod,JacquodRev}, were motivated as suitable approximations of the 
quantum mechanical evolution, in particular, for studies of the different decoherence 
properties between classically regular and chaotic systems. Our results seem 
to show that crossing the quantum-classical divide may offer an additional resource 
for entanglement generation and related ``truly quantum'' phenomena. This might be related to   
a common ``underlying reality'' of CL and QM physics, assumed to consist, for example,  
only in statistical correlations in Refs.\,\cite{Zwitters,Khrennikov}.}    

Concerning the quantum-classical divide, the present 
analysis shows that there is an appealing, if not puzzling formal similarity between CL and 
QM. However, this demonstrates one more time that what has been discussed in various ways 
as CL limit of QM -- and which is similarly relevant for ``emergent QM'' -- deserves further 
study. 

While our work has been concerned mainly with the evolution of CL or QM objects, 
we recall that V.I.\,Man'ko and collaborators have pointed out that classical 
states may differ widely from what could be obtained as the ``$\hbar\rightarrow 0$'' 
limit of quantum mechanical ones. They show that all states can 
be classified by their `tomograms' as {\it either} CL {\it or} QM, CL {\it and} QM, 
and {\it neither} CL {\it nor} QM~\cite{Manko}.  

The classical limit can be considered a limit 
``{\it F}or{\it A}ll{\it P}ractical{\it P}urposes'', gradually approached with decoherence 
as an essential but insufficient ingredient or, formally, following the mnemonic 
``$\hbar\rightarrow 0$'' rule. 
However, in order to truly bridge the qualitative difference between intra- and inter-space 
entanglement that we find, and explain the ``Man'ko classes of states'', 
some unknown dynamics seems missing.~\footnote{A simple attractor model, 
motivated by assumptions about effects of fundamental spacetime discreteness~\cite{Elze09a}, 
has been discussed in Ref.~\cite{Elze09b}.} 

The problems discussed here lead us to the question:   
``Does the $\hbar\rightarrow 0$ deformation of QM provide the only {\it interesting 
linear dynamics} besides QM itself?'' 

\section{How special is quantum mechanics?}
We are interested here once more in the structure of the linear evolution equations 
that we have discussed side by side, namely the classical Liouville equation and 
the von\,Neumann equation of quantum mechanics. However, we consider this with respect 
to a hypothetical most {\it general linear dynamics} that can be represented in the 
generic form of Eq.\,(\ref{linform}): 
\begin{equation}\label{linform1} 
\partial_t\rho_{ij}=\frac{1}{i}\sum_{k,l}{\cal L}_{ij,kl}\rho_{kl} 
\;\;, \end{equation} 
where the indices ($i,j,k,l=1,\dots ,N$) refer to a discrete, finite dimensional 
Hilbert space, which we assume for simplicity. Thus, generally, there are $N^2\times N^2$ 
complex coefficients to be specified. Imposing the {\it constraint} that the `density 
matrix' is and remains Hermitian, which requires that 
${\cal L}_{ij,kl}=-{\cal L}_{ji,lk}^\ast$, the set of coefficients 
can be specified by $N^4$ {\it real parameters}.   
  
This should be compared with the von\,Neumann equation -- which we discussed, in order to 
introduce the concept of a superoperator, cf. Eqs.\,(\ref{vN})--(\ref{LvN}) in Section~2.2. 
In terms of a Hermitian Hamiltonian, $\hat H$, the matrix elements of which are  
specified by $N^2$ {\it real parameters}, the corresponding superoperator has been obtained as:     
${\cal L}_{ij,kl}\equiv H_{ik}\delta_{lj}-\delta_{ik}H_{lj}$.   
  
Therefore, we can write a symbolic relation, concerning   
the number of real parameters which determine the respective dynamical 
equation:
\begin{equation}\label{mnemonic}
``\; \mbox{QM}\sim\sqrt{\mbox{GL}}\; " 
\;\;, \end{equation} 
i.e., the number of real parameters entering the quantum mechanical evolution law (QM)
scales with the Hilbert space dimension like the square root of the corresponding 
number for the most general linear dynamics (GL) which preserves hermiticity. 
By choosing the eigenstates of the Hamiltonian as basis of the Hilbert space,  
the number of relevant parameters in QM could be further reduced to $N$, the number of real 
eigenvalues of $\hat H$. 

We may wonder 
whether the commutator in Eq.\,(\ref{vN}), $[\hat H,.]$, presents  
the {\it minimal structure} preserving hermiticity, normalization, and positivity   
of the density matrix. -- Conversely, can 
the different numbers of real parameters between a more general linear evolution 
law and QM be attributed to an {\it attractor mechanism} or {\it information loss} if and when 
QM is emergent \cite{tHooft10,Elze09a}? 

As a first step to investigate these issues, we consider here whether QM admits a more general 
linear evolution. -- We recall the following result for the QM 
of {\it open systems}, see Ref.\,\cite{Diosi} and references there: 
\begin{itemize}  
\item If and only if the Liouville superoperator $\hat{\cal L}$ entering the right-hand side 
of Eq.\,(\ref{linform1}) generates a 
{\it Hermitian, trace preserving}, {\it completely positive map},   
$\hat{\cal M}(t)\hat\rho (0):=\exp (-i\hat{\cal L}t)\hat\rho (0)$, 
then it can be written in canonical {\it Kraus form}:     
\begin{equation}\label{KrausForm}
\hat{\cal M}(t)\hat\rho (0) =\sum_k\hat M_k(t)\hat\rho (0)\hat M_k^{\;\dagger}(t) 
\;\;, \end{equation}
with $\sum_k\hat M_k^{\;\dagger}\hat M_k={\mathbf 1}_{N\times N}$. 
\item Such a map $\hat{\cal M}$ applied to a  
Hermitian, positive semidefinite, normalized density operator yields another one  
and linear maps beween density operators can be written in Kraus form.  
\end{itemize}
An evolution equation corresponding to such a map, in general, contains the  
von\,Neumann equation as a special case. --   
In fact, the solution of the von\,Neumann equation (\ref{vN}) is 
provided by a unitary transformation, 
$\hat\rho (t)=\hat U(t)\hat\rho (0)\hat U(t)^\dagger$, 
which presents the simplest case with only one unitary Kraus operator, 
$\hat M(t)\equiv \hat U(t):=\exp (-i\hat Ht)$. 

Therefore, any generalization  
which maintains the defining properties of density operators, generally,  
will need additional Kraus operators specified by additonal parameters. -- 
In this sense, the commutator defining the von\,Neumann equation presents 
the {\it minimal structure} indeed.  

Furthermore, 
it has been shown that the dynamics generated by $\hat {\cal M}$ of the Kraus form, 
in general,   
is equivalent to a non-unitary reduced dynamics of a unitary dynamics on a bigger  
(tensor product) Hilbert space ``$S$ystem$\,\otimes E$nvironment'' \cite{Diosi}: 
\begin{equation}\label{KrausDyn} 
\hat {\cal M}\hat\rho_S = 
\mbox{Tr}_E\left (\hat U_{big}\hat\rho_S\otimes\hat\rho_E\hat U_{big}^\dagger\right ) 
\;\;, \end{equation} 
with $\hat U_{big}\hat U_{big}^\dagger =\hat U_{big}^\dagger\hat U_{big}=1$. 
Hence, we learn:  
\begin{itemize}
\item Hermitian, trace preserving, completely positive maps, while generalizing 
the von\,Neumann dynamics, 
are {\it not} general enough to ``leave QM''. They describe the QM of {\it open systems}. 
\end{itemize}
However, this kind of maps does {\it not} 
exhaust the larger set of maps generated by all possible Liouville superoperators. 

We conclude that    
one cannot ``leave QM'' by invoking {\it more general linear dynamics than that  
generated by trace preserving, completely positive maps} without affecting properties and 
interpretation of the states represented by density matrices.}~\footnote{It will be 
interesting to see how (any form of) hybrid dynamics, cf. Section~3., fares in 
this respect.}   

It is useful to recall here (the motivation behind) 
the assumptions made concerning the 
properties and interpretation of density operators. --  
A density operator $\hat\rho$ is required to be {\bf i)} {\it Hermitian}, 
i.e., to have real eigenvalues, in order to qualify as an observable. This 
is needed, in turn, if one requires $\hat\rho$ to be {\bf ii)} {\it positive-semidefinite} 
and {\bf iii)} {\it normalized}. All three properties, together, are necessary 
assumptions for the standard probability interpretation of the eigenvalues of a 
density matrix, according to the Born rule. 

A recurrent theme, when comparing quantum with classical states, for example, 
with the help of the Wigner (function) transform of the density matrix, is the 
appearance of negative eigenvalues or {\it ``negative probabilities''}. 
Numerous attempts have been made to give a satisfactory physical interpretation 
and mathematically consistent definition 
to these, see Refs.\,\cite{Khrennikov,Mueckenheim,Burgin} and further references there. 

Instead of entering this discussion, we look at the arguably simplest 
example of {\it general linear dynamics} in the following. This is obtained by giving 
up the requirement {\bf ii)} above, i.e., by abandoning positivity, and 
by considering a two-dimensional Hilbert space. 

\subsection{General linear $2\times 2$ dynamics: a model} 
A convenient parametrization of the most general Liouville 
superoperator $\hat{\cal L}$ for a two-dimensional state space 
can be written with the help of the Pauli matrices $\vec\sigma$. 
Thus, the (super)matrix elements ${\cal L}_{ij,kl}$, $i,j,k,l=1,2$, are defined by: 
\begin{equation}\label{Ltwobytwo} 
{\cal L}_{ij,kl}:=i{\cal G}_{\mu\nu}
\langle i|a_\mu+\vec b_\mu\cdot\vec\sigma |k\rangle 
\langle l|a_\nu+\vec b_\nu\cdot\vec\sigma |j\rangle 
\;\;, \end{equation} 
where a summation over repeated indices is understood; in this notation 
we have $(\hat{\cal L}\hat\rho )_{ij}={\cal L}_{ij,kl}\rho_{kl}$. Hermiticity requires 
${\cal G}_{\mu\nu}={\cal G}_{\nu\mu}^\ast$. Generally, three vectors 
$\vec b_{\mu =0,1,2}\in {\mathbf R^3}$ are needed. Furthermore, we define: 
\begin{equation}\label{G} 
{\cal G}:=\left ( \begin{array}{ccc} 
1 & 0 & 0 \\ 
0 & g_{11} & g_{12} \\ 
0 & g_{12}^\ast & g_{22} 
\end{array} \right )
\end{equation} 
i.e., where $g$ is a Hermitian $2\times 2$ matrix. This Ansatz saturates the expected 
number of $N^4=2^4=16$ real parameters, since $a_\mu$, $\vec b_\mu$, and $g$ contribute 
3, 9, and 4 parameters, respectively. In order that the map generated by $\hat {\cal L}$ 
be trace preserving, we must have $0\stackrel{!}{=}i\partial_t\mbox{Tr}\hat\rho 
=\mbox{Tr}\hat {\cal L}\hat\rho$, for all $\hat\rho$, which leads to 
${\cal L}_{ii,kl}\stackrel{!}{=}0$. This yields four real constraints, 
\begin{eqnarray}\label{C1} 
a\cdot {\cal G}\cdot a+\vec b\cdot {\cal G}\cdot\vec b&\stackrel{!}{=}&0 
\;\;, \\ [1ex] \label{C2} 
2a\cdot\mbox{Re}{\cal G}\cdot\vec b+ig_{ij}\vec b_i\times\vec b_j&\stackrel{!}{=}&0 
\;\;, \end{eqnarray} 
which reduce the number of available real parameters to twelve. 

Next, we conveniently choose the vectors $\vec b_{0,1,2}$ to form a right-handed 
orthogonal system; with respect to suitable coordinates, this sets six vector 
components to zero. Thus, we are left with six real parameters, three of which 
could characterize the QM evolution of a two-state object, while three pertain 
to the generalization we are concerned with. 

Finally, we assume for simplicity 
that the `QM part' of $\hat {\cal L}$ is diagonal, 
corresponding to a spin-1/2 particle in a constant external magnetic field 
parallel to the quantization axis, for 
example.
Setting one of the remaining parameters to zero, we obtain a simplified model 
with altogether three real parameters, 
$\alpha ,\beta ,\gamma\in\mathbf{R}$: 
\begin{eqnarray}  
{\cal L}_{ij,kl}&:=&\alpha\Big\{\langle i|\sigma_z|k\rangle\delta_{lj} 
-\delta_{ik}\langle l|\sigma_z|j\rangle\Big\} 
\nonumber \\ [1ex] 
&\;&+\beta\gamma\Big\{\langle i|\sigma_z|k\rangle\langle l|\sigma_x|j\rangle
-\langle i|\sigma_x|k\rangle\langle l|\sigma_z|j\rangle\Big\} 
+i\beta\gamma\Big\{\langle i|\sigma_y|k\rangle\delta_{lj} 
+\delta_{ik}\langle l|\sigma_y|j\rangle\Big\} 
\nonumber \\ [1ex] \label{Lsimp}
&\;&+i\beta^2\delta_{ik}\delta_{lj}
+i\gamma^2\langle i|\sigma_y|k\rangle\langle l|\sigma_y|j\rangle
-i(\beta^2+\gamma^2)\langle i|\sigma_z|k\rangle\langle l|\sigma_z|j\rangle
\;\;, \end{eqnarray} 
where QM terms $\propto\alpha\sigma_z$ feature in the first line. 
All other contributions on the right-hand side have no counterpart in 
QM; this will become obvious by explicitly solving the model.
The first two terms in the second line and the two last ones in the third couple 
bra- and ket-states, 
which we discussed as a consequence of classical evolution in 
Sections~2.1 and 3.1. It is easy to verify that 
the generator $-i\hat {\cal L}$ is Hermitian and trace preserving, as it should.

Taking hermiticity and trace normalization of the density matrix into account by: 
\begin{equation}\label{rhoparam} 
\hat\rho\equiv
\left ( \begin{array}{cc} 
\rho_{11} & \rho_{12} \\ 
\rho_{12}^\ast & 1-\rho_{11} 
\end{array} \right )
\;\;, \end{equation} 
we write the resulting evolution equation explicitly: 
\begin{eqnarray}\label{GLevol}   
&\;&i\partial_t\hat\rho\; =\;\hat{\cal L}\hat\rho \; =\;
\\ [1ex] \nonumber 
&\;&\left ( \begin{array}{cc} 
i\gamma^2(1-2\rho_{11}) & 2\beta\gamma +(2\alpha +i[2\beta^2 +\gamma^2])\rho_{12}
-i\gamma^2\rho_{12}^\ast \\  
-2\beta\gamma -(2\alpha -i[2\beta^2 +\gamma^2])\rho_{12}^\ast 
-i\gamma^2\rho_{12} & -i\gamma^2(1-2\rho_{11}) 
\end{array} \right )
. \end{eqnarray} 
Note that for $\beta,\gamma\rightarrow 0$, we recover the von\;Neumann equation, e.g., for 
a spin-1/2 particle in a constant magnetic field. In this case, besides the Hamiltonian, 
$\hat H:=\alpha\sigma_z$, also the 
total spin-squared $S^2:=\vec\sigma^2/4\propto\mathbf{1}$ is a constant of motion.       

\subsection{Constants of motion} 
We remark that if a constant operator, 
$\hat C$, obeys $\mbox{Tr}(\hat C\hat\rho (0))=\mbox{Tr}(\hat C\hat\rho (t))$, 
for all solutions $\hat\rho$ of the general linear evolution equation (\ref{linform1}), 
then $C_{ij}{\cal L}_{ji,kl}=0$, which generalizes the vanishing of the  
commutator $[\hat C,\hat H]_{lk}$ in QM, for constants of motion. This is fulfilled 
for any $\hat C\equiv c\mathbf{1}$, with constant $c$, due to the preservation of the trace 
normalization of the density matrix, incorporated by $\hat{\cal L}_{ii,kl}\stackrel{!}{=}0$. 
However, this raises also the interesting question, whether there exist conserved superoperators, 
$\hat Q$, which can be defined by the vanishing supercommutator: 
\begin{equation}\label{superC} 
[\hat Q,\hat{\cal L}]_{ij,mn}:=Q_{ij,kl}{\cal L}_{kl,mn}-{\cal L}_{ij,kl}Q_{kl,mn}
\stackrel{!}{=}0 
\;\;. \end{equation} 
In the present case, this amounts to a $4\times 4$ matrix equation, to be studied. 

\subsection{Solution of the $2\times 2$ model}
We now turn to the explicit solution of Eq.\,(\ref{GLevol}), which can be represented 
as follows. The matrix elements are: 
\begin{eqnarray}\label{rho11}
\rho_{11}(t)=1-\rho_{22}(t)&=&\frac{1}{2}
+\Big (\rho_{11}(0)-\frac{1}{2}\Big )\mbox{e}^{-2\gamma^2t} 
\;\;, \\ [1ex] \label{rho12}
\rho_{12}(t)=\rho_{21}^\ast (t)&=& 
r_+\mbox{e}^{\Omega_+t}+r_-\mbox{e}^{\Omega_-t}+r_c 
\;\;, \end{eqnarray} 
where $r_{\pm}$ are determined by initial conditions.~\footnote{Interestingly, the 
underlying equation for the off-diagonal matrix elements, 
for $\gamma\neq 0$ in particular, is of second order; 
it decouples into two first order equations in the limit $\gamma\rightarrow 0$, 
independently of the value of $\beta$.}  
The remaining constants 
are defined by: 
\begin{eqnarray}\label{Rc} 
r_c&:=&-\beta\gamma\frac{\alpha -i\beta^2}{\alpha^2+\beta^2(\beta^2+\gamma^2)} 
\;\;, \\ [1ex] \label{Omega}
\Omega_\pm&:=&2\beta^2+\gamma^2
\pm 2i\alpha (1-\gamma^4/4\alpha^2)^{1/2} 
\;\;, \end{eqnarray}
and we may choose $\alpha\geq 0$. 
Diagonalizing the resulting traceless Hermitian density matrix, 
we obtain the eigenvalues: 
\begin{equation}\label{eigenval} 
\rho_\pm (t)=\frac{1}{2}
\pm\Big ((\rho_{11}(0)-1/2)^2\mbox{e}^{-4\gamma^2t}+|\rho_{12}(t)|^2\Big )^{1/2}
\;\;, \end{equation} 
with $|\rho_{12}|^2$ given by Eqs.\,(\ref{rho12})--(\ref{Omega}).  
  
Several features of the solutions of our model seem unexpected. -- 
To begin with, for $\beta,\gamma\rightarrow 0$, we obtain: 
\begin{eqnarray}\label{rho11s}
\rho_{11}(t)=1-\rho_{22}(t)&=&\rho_{11}(0)
\;\;, \\ [1ex] \label{rho12s}
\rho_{12}(t)=\rho_{21}^\ast (t)&=& 
r_+\mbox{e}^{2i\alpha t}+r_-\mbox{e}^{-2i\alpha t}  
\;\;, \end{eqnarray} 
which amounts to the usual QM solution only, if we choose $r_+\equiv 0$. 
From a QM perspective, the {\it larger set of solutions}, parametrized by a 
larger set of initial conditions, is quite surprising. It leads to 
time dependent `probabilities', if we try to interpret the eigenvalues 
of $\hat\rho$ in the usual way: 
\begin{equation}\label{probabs} 
\rho_\pm (t)=\frac{1}{2}
\pm\Big ((\rho_{11}(0)-1/2)^2
+|r_+|^2+|r_-|^2+2\mbox{Re}(r_+r_-^\ast\mbox{e}^{4i\alpha t})\Big )^{1/2}
\;\;. \end{equation} 
While these `probabilities' are real and normalized, by 
construction of our model, they might temporarily fall outside 
the interval $[0,1]$, depending on the parameters $r_\pm$. 

This remark  
applies also for the general case, with $\beta,\gamma\neq 0$, where amplitudes 
grow $\propto\exp (2\beta^2+\gamma^2)t$, see Eqs.\,(\ref{rho12}),(\ref{Omega}). 

The effect of such anomalous `probabilities' is clearly seen in 
the following mean square deviations sensitive to fluctuations. 
With $\langle\hat O\rangle :=\mbox{Tr}(\hat O\hat\rho )$, 
$\sigma_{x,y,z}^2=\mathbf{1}$, and 
$\langle\sigma_x\rangle =2\mbox{Re}(\rho_{12})$, $\langle\sigma_y\rangle =-2\mbox{Im}(\rho_{12})$, 
$\langle\sigma_z\rangle =2\rho_{11}-1$, we obtain: 
\begin{eqnarray}\label{x} 
\Delta_x^2:=
\langle\sigma_x^2\rangle -\langle\sigma_x\rangle^2&=&1-\Big (2\mbox{Re}(\rho_{12})\Big )^2
\;\;, \\ [1ex] \label{y} 
\Delta_y^2:=
\langle\sigma_y^2\rangle -\langle\sigma_y\rangle^2&=&1-\Big (2\mbox{Im}(\rho_{12})\Big )^2 
\;\;, \\ [1ex]\label{z} 
\Delta_z^2:=
\langle\sigma_z^2\rangle -\langle\sigma_z\rangle^2&=&4\rho_{11}(1-\rho_{11})
\;\;. \end{eqnarray} 
While $\Delta_z^2\stackrel{t\rightarrow\infty}{\longrightarrow}1$, for $\gamma\neq 0$, 
which would correspond to a completely mixed state in QM, $\Delta_{x,y}^2$ can have 
oscillatory contributions with growing amplitude  
$\propto\exp\Omega_\pm t$, 
see Eqs.\,(\ref{rho11})--(\ref{Omega}). -- Also note that the would-be 
QM energy expectation varies in time, 
$\alpha\langle\sigma_z\rangle =\alpha (2\rho_{11}(0)-1)\exp (-\gamma^2t)$, approaching 
the stationary mixed-state value zero.   

To summarize: {\it Modifying a QM two-state model  
by applying a general linear perturbation, however small --  
which preserves hermiticity and trace normalization of the 
density matrix --  
opens the possibility of a larger state space, reflected in a  
doubling of the number of degrees of 
freedom}.~\footnote{I.e., a doubling of the initial conditions for the 
off-diagonal matrix elements here; it will be interesting to see, whether this extends 
to all degrees of freedom in a more general model.} 

We emphasize that the QM evolution becomes exponentially unstable when 
such perturbation is introduced. 
Generically, this spoils the standard interpretation of the eigenvalues 
of the density matrix as probabilities and can  
result in interesting oscillatory effects, as we have seen.~\footnote{One might  
speculate on the relevance for flavour oscillations.} 

It is an interesting question, 
whether there is a ``classical'' formulation of general linear dynamics, 
suggested by a doubled number of certain degrees freedom and ``anomalous probabilities'' 
that seem invariably to appear -- and which remind of similar phenomena 
encountered when relating quantum and classical mechanics, e.g., via the  
Wigner function.    

\section{Conclusions}  
In this article we have summarized our earlier derivation of a {\it path integral}  
for classical Hamiltonian systems based on an ensemble description \cite{EGV10,FabThesis}. 

This leads us to point out the correlation 
properties of classical dynamics in parallel to the quantum mechanical 
ones and to identify characteristic similarities and differences. In particular, 
it seems useful to distinguish {\it intra- and interspace entanglement}. 

The former has been held characteristic of quantum mechanics and a feature of 
superpositions of tensor product states. 
The latter concerns classical mechanics only; it correlates  
Hilbert space states and their duals. However, surprisingly, for anharmonic 
potentials or interactions, classical mechanics additionally shows intraspace entanglement, 
as in quantum mechanics.  

As a first application, we propose a new formulation of {\it hybrid dynamics}, 
i.e., based on a hypothetical direct coupling between quantum and classical objects. 
This may be of practical as well as foundational interest. 

Finally, we study a generalization of quantum evolution, 
{\it general linear dynamics}, where the evolution is generated  
by a superoperator that preserves hermiticity and trace normalization of 
density matrices. We argue that one cannot ``leave QM'' without giving up 
one of the three defining properties of density matrices, to be Hermitian,  
normalized, and positive-semidefinite. 

In the most simple case of a two-state system, we solve such dynamics explicitly.  
We show that the corresponding von\,Neumann dynamics becomes 
{\it exponentially unstable} under the influence of a general linear perturbation. 
Most interestingly, it leads to the appearance of ``anomalous probabilities'' 
and an enlargement of the state space, possibly pointing towards a   
sort of prequantum dynamics. -- We intend to study more complex 
objects consisting of such two-state systems as   
building blocks which interact. This may be useful in 
trying to understand quantum mechanics as an emergent phenomenon \cite{tHooft10}.      


\ack{We thank M.J.~Everitt, F.~Finster, A.~Khrennikov, V.I.~Man'ko, and T.~Padmanabhan 
for discussions.}



\section*{References}
\begin{thebibliography}{99}


\bibitem{tHooft10} 't\,Hooft G 2010 Classical cellular automata and quantum field theory 
{\it Int. J. Mod. Phys.} A {\bf 25} No.~23 4385-4396; 
{\it do.} 2009 {\it Preprint} arXiv:0908.3408     

\bibitem{Elze09a} Elze H-T 2009 Does quantum mechanics tell an atomistic spacetime? 
{\it J. Phys.: Conf. Ser.} {\bf 174} 012009 ({\it Preprint} arXiv:0906.1101) 

\bibitem{Elze09b} Elze H-T 2009 {\it Int. J. Qu. Inf. (IJQI)} {\bf 7} 83-96 
({\it Preprint} arXiv:0806.3408)



\bibitem{Elze08} Elze H-T 2008 {\it J. Phys. A.: Math. Theor.} {\bf 41} 304020  
({\it Preprint} arXiv:0710.2765) 

\bibitem{tHooft07} 't\,Hooft G 2007 {\it AIP Conf. Proc.} {\bf 957} 154-163 
({\it Preprint} arXiv:0707.4568)    

\bibitem{tHooft06a} 't\,Hooft G 2007 {\it J. Phys.: Conf. Ser.} {\bf 67} 012015   
({\it Preprint} arXiv:quant-ph/0604008)

\bibitem{tHooft06b} 't\,Hooft G 2003 {\it Int. J. Theor. Phys.} {\bf 42} 355-361; 
{\it do.} 1999 {\it Class. Quant. Grav.} {\bf 16} 3263-3279  

\bibitem{Elze05} Elze H-T 2006 {\it J. Phys.: Conf. Ser.} {\bf 33} 399-404 
({\it Preprint} arXiv:gr-qc/0512016); 
{\it do.} 2005 {\it Braz. J. Phys.} {\bf 35} 343-350; 
{\it do.} 2003 {\it Phys. Lett.} A {\bf 310} 110-118  

\bibitem{Blasone05} Blasone M, Jizba P and Kleinert H 2005 {\it Ann. Phys.} {\bf 320} 468-486 
({\it Preprint} arXiv:quant-ph/0504200); 
{\it do.} 2005 {\it Braz. J. Phys.} {\bf 35} 497-502; 
{\it do.} 2005 {\it Phys. Rev.} A {\bf 71}  052507  

\bibitem{Vitiello01} Blasone M, Jizba P and Vitiello G 2001 
{\it Phys. Lett.} A {\bf 287} 205-210 ({\it Preprint} arXiv:hep-th/0007138)  

\bibitem{Smolin} Markopoulou F and Smolin L 2004 {\it Phys. Rev.} D {\bf 70} 124029 
({\it Preprint} arXiv:gr-qc/0311059)  

\bibitem{Adler} Adler S L 2005 {\it Quantum Mechanics as an Emergent Phenomenon} 
(Cambridge, UK: Cambridge U. Press) 

\bibitem{Wetterich08} Wetterich C 2008 {\it J. Phys.: Conf. Ser.} {\bf 174} 012008 
({\it Preprint} arXiv:0811.0927); 
{\it do.} 2008 Probabilistic observables, conditional correlations, and quantum physics 
{\it Preprint} arXiv:0810.0985; 
{\it do.} 2008 Quantum entanglement and interference from classical statistics 
{\it Preprint} arXiv:0809.2671

\bibitem{Isidro08} Isidro J M, Santander J L G and Fernandez de Cordoba P 2008 
A note on the quantum-mechanical Ricci flow {\it Preprint} arXiv:0808.2717;
{\it do.} 2008 Ricci flow, quantum mechanics and gravity {\it Preprint} arXiv:0808.2351 

\bibitem{Iwo} Bialynicki-Birula I and Mycielski J 1976 {\it Ann. Phys. (N.Y.)} {\bf 100} 62   

\bibitem{Kibble78} Kibble T W B 1978 {\it Commun. Math. Phys.} {\bf 64} 73  

\bibitem{Kibble80} Kibble T W B and Randjbar-Daemi S 1980 {\it J. Phys.} A {\bf 13} 141  

\bibitem{Weinberg} Weinberg S 1989 {\it Phys. Rev. Lett.} {\bf 62} 485; {\it do.} 1989 
{\it Ann. Phys. (N.Y.)} {\bf 194} 336 

\bibitem{L1} Lindblad G 1976 {\it Commun. Math. Phys.} {\bf 48} 119-130 

\bibitem{L2} Gorini V, Kossakowski A and Sudarshan E C G 1976 {\it J. Math. Phys.} 
{\bf 17} 821-825  

\bibitem{Huetal} Chou C H, Hu B-L and Suba\c{s}i Y 2011 Macroscopic quantum phenomena 
from the large N perspective {\it Preprint}, in this proceedings volume 

\bibitem{ZhangWu06} Zhang Q and Wu B 2006 {\it Phys. Rev. Lett.} {\bf 97} 190401 

\bibitem{Diosi10} Di\'osi L 2010 The gravity-related decoherence master equation from 
hybrid dynamics {\it Preprint} arXiv:1101.0672, in this proceedings volume  

\bibitem{Diosi84} Di\'osi L 1984 {\it Phys. Lett.} A {\bf 105} 199   

\bibitem{DiosiRev05} Di\'osi L 2005 {\it Braz. J. Phys.} {\bf 35} 260-265 
({\it Preprint} arXiv:quant-ph/0412154v2) 

\bibitem{Penrose98} Penrose R 1998 {\it Phil. Trans. R. Soc.} {\bf 356} 1927  

\bibitem{Adler03} Adler S L 2003 {\it Stud. Hist. Philos. Mod. Phys.} {\bf 34} 135  

\bibitem{Mavromatosetal92} Ellis J, Mavromatos N E and Nanopoulos D V 1992 
{\it Phys. Lett.} B {\bf 293} 37-48 ({\it Preprint} arXiv:hep-th/9207103v2)   

\bibitem{Pullinetal08} Gambini R, Porto R A and Pullin J 2008 
{\it Phys. Lett.} A {\bf 372} 1213-1218 ({\it Preprint} arXiv:0708.2935v2) 

\bibitem{Hu09} Hu B-L 2009 {\it J. Phys.: Conf. Ser.} {\bf 174} 012015 
({\it Preprint} arXiv:0903.0878)   

\bibitem{Diosi09} Di\'osi L 2009 {\it J. Phys.: Conf. Ser.} {\bf 174} 
({\it Preprint} arXiv:0902.1464) 012002   

\bibitem{EGV10} Elze H-T, Gambarotta G and Vallone F 2011 A path integral for classical dynamics, 
entanglement, and Jaynes-Cummings model at the quantum-classical divide 
{\it Int. J. Qu. Inf. (IJQI)}, in press ({\it Preprint} arXiv:1006.1569)   

\bibitem{Elze07} Elze H-T 2007 {\it Int. J. Theor. Phys.} {\bf 46} No.~8 2063-2081 
({\it Preprint} arXiv:hep-th/0510267)   

\bibitem{superspace} Mukamel S 1995 {\it Principles of nonlinear optical spectroscopy} 
(Oxford: Oxford Univ. Press) 

\bibitem{Schulman} Schulman L S 1981 {\it Techniques and applications of path integration} 
(New York: Wiley)  

\bibitem{Kleinert} Kleinert H 2009 {\it Path integrals in quantum mechanics, statistics, 
polymer physics, and financial markets} 5th edition (Singapore: World Scientific) 

\bibitem{FabThesis} Vallone F 2010 {\it } (Master thesis, Universit\'a di Pisa), unpublished  

\bibitem{Jacquod} Jacquod Ph 2004 {\it Phys. Rev. Lett.} {\bf 92} 150403  
({\it Preprint} arXiv:quant-ph/0308099)  

\bibitem{JacquodRev} Jacquod Ph and Petitjean C 2009 
{\it Adv. in Phys.} {\bf 58} vol. 2, 67 ({\it Preprint} arXiv:0806.0987)   

\bibitem{Hornberger} Busse M and Hornberger K 2010 
{\it J. Phys. A: Math. Theor.} {\bf 43} 015303 ({\it Preprint} arXiv:0910.1062) 

\bibitem{Zwitters} Wetterich C 2009 Zwitters: particles between quantum and classical  
{\it Preprint} arXiv:0911.1261 

\bibitem{Khrennikov} Khrennikov A 2003 {\it ``Interpretations of Probability''} 
(Utrecht and Boston: VSP) 

\bibitem{Manko} Man'ko O V and Man'ko V I 2004 {\it J. Russ. Laser Res.} {\bf 25}(5) 
477 ({\it Preprint} arXiv:quant-ph/0407183)   

\bibitem{Diosi} Di\'osi L 2007 {\it A Short Course in Quantum Information Theory, 
Lecture Notes in Physics} {\bf 713} (Berlin: Springer) 
 
\bibitem{Mueckenheim} M\"uckenheim W 1986 {\it Phys. Rep.} {\bf 133}(6) 337   

\bibitem{Burgin} Burgin M 2010 Interpretations of Negative Probabilities 
{\it Preprint} arXiv:1008.1287 





















\end{thebibliography}
\end{document}